\documentstyle[12pt]{article}
\date{}
\begin{document}

{\large\rm DESY 94-046}\hfill{\large\tt ISSN 0418-9833}

{\large\rm March 1994}\hfill\vspace*{3.5cm} \hfill
\begin{center}
{\large {\bf Overrelaxation Algorithm for coupled\\
Gauge-Higgs Systems}}\\
\vspace{0.9truecm}
Zolt\'an Fodor\footnote{On leave from
Institute for Theoretical Physics, E\"otv\"os
University, Budapest, Hungary} and Karl Jansen\\
\normalsize\it  Deutsches Elektronen-Synchrotron DESY, Hamburg, Germany
\end{center}
\sloppy
\setlength{\baselineskip}{18pt}
\vspace*{3cm}
\begin{abstract}
In this letter we extent the overrelaxation algorithm,
known to be very efficient in gauge theories,
to coupled
gauge-Higgs systems with a particular emphasis on the update of
the radial mode of the Higgs field. Our numerical tests
of the algorithm show that the
autocorrelation times can be reduced substantially.
\end{abstract}
\vfill\eject

\section{}
The electroweak baryogenesis has attracted a lot of attention recently
\cite{kaplanrev}.
In order to understand this phenomenon a detailed description of
the electroweak phase transition (PT) is needed.
In its study it is sufficient
to concentrate on the SU(2) group alone
and to neglect all fermion fields and the U(1)
subgroup.
One therefore remains
with a coupled system of gauge and Higgs fields
which are elements of SU(2) and complex doublets, respectively.
Although this so-called SU(2)-Higgs model has been investigated up
to 2-loop order at finite temperature
in perturbation theory by now \cite{fodor},
numerical
simulations, which intrinsically contain also non-perturbative effects, are
desirable. This is even more so as the 2-loop results showed large
corrections to quantities like the surface tension as
compared to 1-loop results.
Therefore it seems necessary to confront the perturbative approach with
numerical ``experiments'' to test the reliability of the perturbative
results.

The Monte-Carlo simulations are necessarily done on a
discretized (euclidean) space time lattice. To obtain results which are
not distorted by the lattice one would like to go close to the continuum
and work in the "scaling region".
To approach this limit
some of the correlation lengths $\xi_i$ have to diverge.
This causes the basic problem of all numerical simulations as with growing
correlation lengths the
problem of critical slowing
down arises. In order to obtain independent configurations, suitable for
measurements of physical quantities, one has to take into account the
autocorrelation time which is the number of iterations with a
given algorithm to reach a new independent configuration. The autocorrelation
time grows with the correlation length as $\tau \propto \xi^z$ with $z$ the
so called dynamical critical exponent. For local algorithms like
Metropolis or heatbath $z$ is known to be 2. This implies that
for $\xi$'s of the order of 5-10, which are realistic values of todays
simulations of the SU(2)-Higgs model,
the autocorrelation time can be of the order
of 25-100. Indeed, as will be shown below, in the SU(2)-Higgs model, for
parameter values where one can compare
results from perturbation theory and numerical
simulations, the autocorrelation times are $O(100)$.  This appears to be
a major drawback for Monte-Carlo studies of the electroweak PT.

It is consequently not surprising that fighting critical slowing down is
one of the major activities in the area of Lattice field theory. In fact,
in the last few years several important steps have been
done to solve this problem. Cluster \cite{wolffrev} and multigrid
\cite{multigridrev} techniques
are able to reduce critical slowing down almost completely, giving
$z \approx 0$. However, these techniques are so far either only applicable for
spin models (cluster) or did not lead to a big improvement in
non-abelian gauge theories (multigrid).

For gauge theories
another interesting approach, the
overrelaxation method, as initiated by Adler \cite{adler} as a generalization
of the heatbath algorithm seems to be most promising.
In its popular limiting case the field
evolution is deterministic and conserves energy (microcanonical).
In order to restore the necessary ergodicity one uses standard,
ergodic updates (e.g. Metropolis or heatbath) and microcanonical
updates alternately
with a given mixing ratio. These so called hybrid algorithms turned
out to be extremely efficient in pure gauge theories \cite{puregauge}. In
this case the overrelaxation step is a kind of reflection of
the local field with the following features: \hfill\break
a) the energy is unchanged \hfill\break
b) since the reflection is an element of the gauge group,
the measure in the defining functional integral is invariant
\hfill\break

\section{}
We will extent this approach to coupled gauge-Higgs systems.
Consider an SU(2) gauge-Higgs system with the following lattice
action:
\begin{equation}
S=S_g +S_h,
\end{equation}
\noindent where
\begin{equation}
S_g =-\frac{\beta}{2}  \mbox{Tr}(\sum_p U_p)
\end{equation}
is the usual Wilson plaquette action and
\begin{equation}
S_h=-\kappa \mbox{Tr}(\sum_{x,\mu}\Phi_x^{\dag} U_{x,\mu}\Phi_{x+\mu})
+\sum_x \left[ \Phi_x^{\dag}\Phi_x +\lambda (\Phi_x^{\dag}\Phi_x -1)^2 \right]
\end{equation}
describes the self interaction of the four component Higgs
field $\Phi_x$ and its coupling to the gauge field. Here one
can introduce the
standard $2 \times 2$ matrix notation, $\Phi_x=\rho_x \alpha_x$,
where $\rho_x \in R$, $\rho_x\ge 0$,  and $\alpha_x \in SU(2)$.

The reflection of the gauge field and the angular part of the
Higgs field can be done analogously as for pure gauge
fields and the above a) and b) conditions can be satisfied.
However, for the radial part of the scalar field the reflection
fulfills a) but not b), thus a careful treatment of
the measure is needed in order to obtain a proper technique.

First we will study the microcanonical updating of the link
variables. The action can be written as a sum of two terms:
a term which contains the local link variable and a constant term.
\begin{equation}
S = \mbox{Tr} (  U_{x,\mu} V)+const,
\end{equation}
where
\begin{equation}
V=-\frac{\beta}{2} \sum_{\nu \neq \mu, \nu \neq -\mu}
U_{x+\mu,\nu} U_{x+\nu,\mu}^{\dag}  U_{x,\nu}^{\dag}
-\kappa \sum_\mu \rho_x \rho_{x+\mu} \alpha_{x+\mu} \alpha_x^{\dag} .
\end{equation}
Consider an update as a ``reflection'' of $U$
\begin{equation}
U'_{x,\mu}=V_0^{\dag} U_{x,\mu}^{\dag} V_0^{\dag},
\end{equation}
where $V_0$ is the normalized $V$, thus $qV_0=V$, where
$q\in R$ and $ V_0 \in SU(2)$. (Note the special feature
of the $SU(2)$ group that the sum of several
$SU(2)$ matrices with real coefficients is a product of a real number and an
$SU(2)$ matrix.)
As in the case of pure gauge theory it is easy to see
that this updating satisfies the above mentioned a) and b)
conditions.

The microcanonical updating of the angular part of the scalar field
can be done completely analogously.
\begin{equation}
S=Tr(\alpha_x^{\dag} V)+const,
\end{equation}
where
\begin{equation}
V=-\kappa \rho_x \sum_\mu \rho_{x+\mu}U_{x,\mu}\alpha_{x+\mu},
\end{equation}
thus the updated angle
\begin{equation}
\alpha_x'=V_0\alpha_x^{\dag} V_0,
\end{equation}
where $V_0$ is, as above, the normalized $V$. Again, one can show that
the above a) and b) conditions are satisfied.

The most interesting case is the microcanonical updating of the radial mode
of the scalar field. Separating the angular and the radial modes as
$\Phi=\rho\alpha$, the ${\cal D}\Phi{\cal D}\Phi^\dagger$
measure will get a form of
$d\alpha d\rho \rho^3$. Therefore one gets a potential
\begin{equation}
V(\rho_x)=-C\rho_x+\rho_x^2+\lambda(\rho_x^2-1)^2-3log(\rho_x),
\end{equation}
where
\begin{equation}
C=\kappa \mbox{Tr}(\sum_\mu \rho_{x+\mu} \alpha_x^{\dag} U_{x,\mu}
\alpha_{x+\mu})\;\;.
\end{equation}

We have plotted this local $V(\rho)$ potential in fig. 1. for
$\lambda=0.0001$ and for a typical $C=2$ value.
The potential has one minimum, but is not symmetric with respect to
its minimum. By a simple analysis of $V(\rho)$ one can show, that it has
only one minimum for $\lambda<(3/2+\sqrt{11}/2)^2\approx9.98$.
For even larger values of $\lambda$, one needs $\kappa\approx {\cal O} (10 )$
to have a second minimum. These $\kappa$-s are clearly out of any
range of physical interest. Thus, we can assume that $V(\rho)$ has
only one minimum.

The procedure to perform the overrelaxation step in $\rho$ is now clear.
For each $\rho_x$ calculate the value of $\rho_x^{'}\ne\rho_x$
such that $V(\rho_x) =
V(\rho_x^{'})$. Note that this is not just a reflection with respect to the
minimum of the potential as $V(\rho)$ is not symmetric.
In practice we found that a reliable way to determine $\rho^{'}$ is to
calculate the first, second and third derivatives of $V$ (analytically)
and calculate $\rho_x^{'}$ for which the Taylor polynom gives
the same value for the potential. After this starting step one-two
Newton iterations give the proper $\rho_x^{'}$ value with a relative
accuracy of $10^{-6}$. If the first step can not be made, or
turns out to be too rude Newton alone or bisection
can be used.

The non-trivial part of the overrelaxation step in $\rho$ is the observation
that $\rho^{'}$ is a nonlinear function of $\rho$.
Therefore, the transformation of $\rho$ to $\rho^{'}$ does not leave
the measure invariant.
As a consequence, one has to correct for this by a reject/accept step.
In order to obtain the proper updating procedure let us consider the
stability equation for $\rho$.

\begin{equation}
\int d\rho T(\rho,\bar \rho)P(\rho)=P(\bar \rho),
\end{equation}
where $P(\rho)$ is the equilibrium distribution in $\rho$ and
$T(\rho,\bar \rho)$ is the overrelaxation operator. It has the form

\begin{equation}
\int_a^bd \rho T(\rho,\bar \rho)= \left\{\begin{array}{lll}
                   0 & \mbox{if} & M(\bar \rho)\notin (a,b)\\
        A(\bar \rho) & \mbox{if} & M(\bar \rho)\in  (a,b)
                   \end{array} \right.
\end{equation}
where $M(\rho)$ is the ``mirror image'' of $\rho$, thus
\begin{equation}
V(M(\rho))=V(\rho) \qquad, \qquad \mbox{but} \qquad \rho \neq M(\rho),
\end{equation}
\noindent unless $\rho$ is exactly at the minimum, where $\rho=$ $M(\rho)$,
of course.
Evaluating the above stability integral on gets
\begin{equation}
\int d \rho A(\bar \rho) \delta(\rho-M(\bar \rho)) P(\rho)=A(\bar \rho)
{1 \over \vert dM(\bar \rho)/d \bar \rho \vert} P(\bar \rho).
\end{equation}
Combining this with the stability equation one obtains
\begin{equation}
A(\rho)=\vert M^{'}(\rho)\vert=\vert {d\rho^{'} \over d\rho} \vert=
\vert {dV \over d\rho}/{dV \over d\rho^{'}}  \vert\;\;.
\end{equation}
The Monte-Carlo realization of the above $T(\rho,\rho^{'})$ overrelaxation
operator can be done in the following way. One determines $M(\rho)$
``mirror image'' of $\rho$ and calculates the $V^{'}(\rho)$ and
 $V^{'}(M(\rho))$ derivatives. The $\rho \rightarrow M(\rho)$ updating
is accepted if
\begin{equation}\label{accept}
\vert {V^{'}(\rho) \over  V^{'}(M(\rho))} \vert > r,
\end{equation}
where $r$ is a random number between $0$ and $1$. Otherwise the
updating is rejected. In practice we found the acceptance always to be
larger than $80\%$.

Note, that the same result can be obtained by considering
the probability densities. The probability of finding the
system in a state with local energy between $V$ and $V+\Delta V$
is a sum of two probabilities. The probability to find it
between $\rho$ and $\rho+\Delta \rho$ plus the probability to find it
between $M(\rho)$ and $M(\rho)+\Delta M(\rho)$, where
$V(\rho)=V(M(\rho))=V$ and
$V(\rho+\Delta \rho)=V(M(\rho)+\Delta M(\rho))=V+\Delta V$.
Since we want the algorithm to be microcanonical,
$dV=V^{'}(\rho)d\rho - V^{'}(\rho^{'})d\rho^{'} =0$ ($\rho^{'}=M(\rho)$).
Clearly one regains (16) for $A(\rho)$ and the reject/accept step (17).

\section{}
We have tested the algorithm described above in the SU(2)-Higgs model.
For parameter values $\lambda=0.0001$, $\beta=8$ and $\kappa=0.129$ the system
is in the Higgs region of the model.
We measured the normalized
autocorrelation function $\Gamma_O$ of some operator $O$
which is defined as
\begin{equation}
\Gamma_O(t) =\frac{<O(0)O(t)> - <O(0)>^2}{<O(0)>^2}
\end{equation}
\noindent where $t$ indicates a fictitious Monte Carlo time corresponding
to the number of sweeps.
$\Gamma_O$ falls off exponentially
\begin{equation}
\Gamma_O(t) \propto \exp(-t/\tau) \;\;\mbox{for}\;\; t\rightarrow\infty
\end{equation}
\noindent which defines the exponential autocorrelation time $\tau$.
In fig.2
we plot the autocorrelation function
$\Gamma_\rho$ for the length of the Higgs field $\rho$. This operator
in general shows the longest autocorrelation time in the SU(2)-Higgs model.
The solid line corresponds to a pure Metropolis simulation and the dashed
line shows the result of a hybrid overrelaxation with a mixing ratio
of Metropolis to overrelaxation 1:1.
The measurements have been performed after each sweep of a Metropolis or
an overrelaxation update through the lattice.
Although the lattice used is quite small ($8^4$) the autocorrelation
function for the Metropolis algorithm
shows a very slow fall off, indicating an autocorrelation time of about
$\tau_{Met}\approx 130$.
On the other hand the hybrid overrelaxation algorithm
gives a considerable improvement,
$\tau_{OR}\approx 25$, over the Metropolis
algorithm \footnote{Based on the ideas
of the present letter, a hybrid of heatbath and
overrelaxation was analyzed in \cite{bunk} and
a similar improvement was found.}.
The results for the exponential autocorrelation
times $\tau$ eq.(19), given above, have been checked against
the integrated autocorrelation times and a complete agreement has been
found.

The dashed curve in fig.2 corresponds to a combination of only one
Metropolis to one overrelaxation step.
We want to note that the main effect of the improvement
stems from the overrelaxation in $\rho$ alone. Switching off the overrelaxation
in $U$ and $\alpha$ we found an almost identical autocorrelation function.
This clearly demonstrates that the operator $\rho$ gives the slowest mode.
Increasing the mixing ratio of Metropolis to overrelaxation steps does not
improve the autocorrelation time. This is understandable from the
$\rho$-potential eq.(10) as more overrelaxation steps reflects $\rho$ only
back and forth, not leading to substantial changes in the configuration.

\section{}

In simulations of the SU(2)-Higgs model, which is
the most relevant part for studies of the
electroweak phase transition, using a Metropolis algorithm one
detects large autocorrelation times already on small lattices.
In order to be able to have reliable numerical results it is therefore
necessary to find an algorithm which improves this behaviour.
We have extended the overrelaxation algorithm which seems to work
very efficient
in pure gauge theories to the SU(2)-Higgs model. The treatment of the
radial part of the Higgs-field in the algorithm led to an additional
accept/reject step which corrects for the measure term.

We have tested our version of the overrelaxation algorithm in the SU(2)-Higgs
model at parameter values interesting for comparisons with perturbation
theory. As fig.2 shows, the autocorrelation time can be reduced
substantially, namely from $\tau \approx 130$ in the Metropolis algorithm
(solid line) to $\tau \approx 25$ in the overrelaxation algorithm.
We want to emphasize that this result is obtained on a small ($8^4$)
lattice. It is expected that the hybrid overrelaxation algorithm
has a dynamical critical exponent of $z\approx 1$ as compared to
the Metropolis or heatbath with $z\approx 2$. Therefore one can
expect even better improvements on larger lattices.

We found it sufficient to perform only one overrelaxation step in the
length of the Higgs field $\rho$ to obtain the best improvement.
We expect, however, that as in pure gauge theories
\cite{puregauge,wolffrev}, one can obtain even smaller autocorrelation
times by tuning the mixing ratio of Metropolis to overrelaxation steps
for the update of the gauge field $U$ and the angle of the Higgs field
$\alpha$. This tuning will depend on the parameter values where simulations
are performed. The optimal mixing ratio should be determined from case to
case.

We think that our numerical results are very promising and
that the hybrid overrelaxation we are suggesting is a big improvement
over algorithms used so far for simulations of the SU(2)-Higgs model.
We hope that by careful fine tuning of the mixing ratio of
different algorithms  for the update of the variables $U$, $\rho$
and $\alpha$
the autocorrelation times might be reduced to small numbers
$\tau < 10$
as it can be obtained in pure gauge theories and one can
therefore obtain high precision numerical data to shed new light on the
the electroweak phase transition beyond perturbation theory.

\section*{Acknowledgment}
We would like to thank B. Bunk, J. Hein,
M. L\"uscher and I. Montvay for helpful
discussion and suggestions.
Z.F. was partially supported by Hung. Sci. Grant under Contract No.
OTKA-F1041/3-2190.

\section*{Figure Caption}

{\bf Fig.1} The local potential eq.(10)
as a function of the radial length of the
scalar field $\rho$ for $\lambda=0.0001$ and $C=2$.

\noindent {\bf Fig.2}
The logarithm of the autocorrelation function for the length of
the Higgs field $\rho$ as a function of the Monte Carlo ``time'' t.
The parameter values are $\beta=8$, $\lambda=0.0001$
and $\kappa=0.129$ and belong to a point in the Higgs region of the model. The
lattice is $8^4$. The solid line is the Metropolis algorithm alone. The dashed
line is a hybrid of one Metropolis to one
overrelaxation step.


\begin{thebibliography}{99}
\bibitem{kaplanrev} A.G. Cohen, D.B. Kaplan and A.E. Nelson, Annu.Rev.Nucl.
                    Part.Sci. {\bf 43} (1993) 27.
\bibitem{fodor} P. Arnold and O. Espinosa, Phys.Rev.D47 (1993) 3546;
                Z. Fodor and A. Hebecker, ``Finite Temperature Effective
                Potential to Order $g^4$, $\lambda^2$ and the
                Electroweak Phase Transition'',DESY preprint, DESY 94-025.
\bibitem{wolffrev} U. Wolff, ``High Precision Simulations with
                   Fast Algorithms'',
                   in Schladming 1992 proceedings (1992) 127.
\bibitem{multigridrev} J. Goodman and A.D. Sokal, Phys.Rev.D40 (1989) 2035.
\bibitem{adler} S.L. Adler, Phys.Rev.D23 (1981) 2901; Phys.Rev.D37
                (1988) 458.
\bibitem{puregauge} K. Decker and Ph. de Forcrand, Nucl.Phys.B
                    (Proc.Suppl.) {\bf 17} (1990) 567.
\bibitem{wolffgauss} U. Wolff, ``Dynamics of Hybrid Overrelaxation in the
                     Gaussian Model", CERN preprint, CERN-TH 6408/92.
\bibitem{bunk} B. Bunk, private communication.
\end{thebibliography}
\end{document}